\documentclass[twocolumn,showpacs,prd,floatfix,axodraw]{revtex4}
\usepackage{mathrsfs}
\usepackage{graphicx,,booktabs,bm}
\usepackage{overpic}
\usepackage{color}
\usepackage{amssymb}

\usepackage{mathrsfs,bm,amsmath,amssymb}
\usepackage{longtable,lscape}
\usepackage{txfonts}
\usepackage{amssymb}
\usepackage{indentfirst}
\usepackage{graphicx,,booktabs}
\usepackage{multirow,ulem}

\usepackage{color}
\usepackage{amssymb}

\definecolor{cover}{rgb}{0.77,0.87,0.88}
\definecolor{blueone}{rgb}{0.1,0.1,.7}
\definecolor{citec}{rgb}{0.14,0.47,0.09}
\definecolor{two}{rgb}{0.0,0.5,0.}
\definecolor{three}{rgb}{.5,.1,0.15}
\usepackage[bookmarks=true,bookmarksopen=false,plainpages=false,breaklinks=true,
   bookmarksnumbered=true,hypertexnames=false,
   filecolor=blue,urlcolor=three,menucolor=three,
   linkcolor=three,citecolor=blueone, colorlinks,
   anchorcolor=blue,runcolor=pink,frenchlinks=red
   pdfstartview=FitH,pdftitle=title,%
   pdfauthor=author]{hyperref}

\begin{document}
\title{Analysis of the $\eta_1(1855)$ as a $K\bar{K}_1(1400)$ molecular state}

\author{Feng Yang}
\affiliation{ School of Physical Science and Technology, Southwest Jiaotong University, Chengdu 610031,China}

\author{Hong Qiang Zhu}
\email{20132013@cqnu.edu.cn}
\affiliation{College of Physics and Electronic Engineering, Chongqing Normal University, Chongqing 401331,China}

\author{Yin Huang\footnote{corresponding author}} \email{yin.huang@apctp.org}
\affiliation{Asia Pacific Center for Theoretical Physics,
Pohang University of Science and Technology, Pohang 37673, Gyeongsangbuk-do,
South Korea}
\affiliation{School of Physical Science and Technology, Southwest Jiaotong University, Chengdu 610031,China}

\begin{abstract}
In this work, we study the radiative and strong decay of $S$-wave $K\bar{K}_1(1400)$ molecular state within the
effective Lagrangians approach and find the relation between the $K\bar{K}_1(1400)$ molecular state and the newly
observed $\eta_1(1855)$ state by comparing with the BESIII observation.   The prediction indicates that the decay
width can reach up to $195.06^{+5.18}_{-5.13}$ MeV, which can be confronted with the
experimental data.  If the $\eta_1(1855)$ could be $S$-wave $K\bar{K}_1(1400)$ molecular state, the $K\bar{K}^{*}\pi$
three-body decay provides the dominant contribution, not the $\eta\eta^{'}$ channel found in the experiment.  In addition,
the partial width for $\eta_1(1855)\to\gamma{}\phi$ can reach up to  $17.67^{+0.38}_{-0.62}$   KeV.  Those results can
be measured in future experiments and used to test the nature of the $\eta_1(1855)$.
\end{abstract}

\date{\today}


\maketitle
\section{Introduction}
In fact, the molecular state composed of $K^{(*)}$ meson in the light flavor sector has been widely studied.   For example,
the $\Lambda(1405)$ seems more natural to be interpreted as a molecular state of $\bar{K}N$
and its coupled channel~\cite{Oset:1997it,Oller:2000fj}, which is supported by the lattice-QCD simulations~\cite{Nemoto:2003ft, Hall:2014uca}.
The structure and quark content of $f_0(980)$ and $a_0(980)$ are predictions as being a $K\bar{K}$ molecule
state~\cite{Ahmed:2020kmp,Dai:2014zta}.  In Refs.~\cite{Roca:2005nm,Geng:2015yta,Aceti:2015zva,Zhou:2014ila,Lutz:2003fm},
the axial-vector meson $f_1(1285)$ can be well taken as a $K^{*}\bar{K}$ molecular state.  Among them, the theoretical calculations on the decay
$f_1(1285)\to \pi{}a_0(980)$ within the $K^{*}\bar{K}$ molecular picture for $f_1(1285)$~\cite{Aceti:2015zva}
have been confirmed in a BESIII experiment~\cite{BESIII:2015you}.  Following the LHCb observation of hidden-charm pentaquark
$P_c(4312,4440,4450)$~\cite{Aaij:2019vzc,LHCb:2020jpq} and their interpretation as $\Sigma^{(*)}_c\bar{D}^{(*)}$~\cite{He:2015cea,Xiao:2019mvs}
molecules, two nucleon resonances with a mass about 2.0 GeV,  the $N(1875)$ and the $N(2100)$, were also interpreted as
hadronic molecular states from the $\Sigma^{*}K$ and $\Sigma{}K^{*}$ interactions, respectively~\cite{He:2017aps,He:2015yva}.
Moreover, the interaction between $\bar{K}^{*}$ and $\Sigma$ can form a $P$-wave molecular state that can be associated to
the $\Xi(2030)$~\cite{Huang:2018uox}.  The narrower width of $\Xi(1620)$, $\Xi(1690)$, and $\Xi(2120)$ can be easy understood
as molecular state with dominant $\bar{K}\Lambda-\bar{K}\Sigma$ component~\cite{Khemchandani:2016ftn,Huang:2020taj,Huang:2021ahp}.
We also note that the $\bar{K}\Xi(1530)$ hadronic molecular picture plays an important role in understanding the observed $\Omega(2012)$~\cite{Huang:2018wth,Valderrama:2018bmv,Lin:2018nqd,Ikeno:2020vqv}p.  The more information about the molecular state
including the $K^{(*)}$ meson can be found in Ref.~\cite{Guo:2017jvc}.

p
Very recently, a meson named $\eta_1(1855)$ was observed by the BESIII Collaboration in the analysis of the
$J/\psi\to\gamma\eta\eta^{'}$ reaction~\cite{Ablikim:2022zze,Ablikim:2022glj}.   The observed resonance masses, widths,
and favorable quantum numbers are
\begin{align}
M&=1855\pm{}9^{+6}_{-1}~~~~~{\rm MeV},\nonumber\\
\Gamma&=188\pm18^{+3}_{-8}~~~~~ {\rm MeV},~~J^{PC}=1^{-+},
\end{align}
respectively.  From the point of the conventional quark states that mesons are made of quark-antiquark pairs and baryons are
composed of three quarks, the spin-parity quantum number of $\eta_1(1855)$ cannot be reproduced.
Hence, from the observed $\eta\eta^{'}$ decay mode.
Since the mass of $\eta(1855)$ is about 40 MeV below the threshold of $K\bar{K}_1(1400)$, it is reasonable to regard it
as a bound state of $K\bar{K}_1(1400)$.  Indeed, the interaction between the $K$ and $\bar{K}_1(1400)$ meson is studied
in the one-boson exchange model and, with reasonable parameters, the $\eta_1(1855)$ can be understood as a $K\bar{K}_1(1400)$
molecule~\cite{Dong:2022cuw}.

Although the $K\bar{K}_1(1400)$ molecular of $\eta_1$(1835) has been successfully explained theoretically, the internal structure
of $\eta_1(1855)$ is still controversial.  The QCD axial anomaly supports the interpretation of the $\eta_1(1855)$ as the $\bar{s}sg$
hybrid meson~\cite{Chen:2022qpd}.  Their results indicate that $\eta\eta^{'}$ decay mode to be a characteristic signal of
the hybrid nature of the $\eta_1(1855)$.  To better understand the binding mechanisms present in multiquark systems and help improve
the understanding of $\eta_1(1855)$, more studies on its spectroscopy and decay width are needed.

In this work, we compute the possible two-body and three-body partial decay widths of $\eta_1(1855)$ by assuming $\eta_1(1855)$ as a
$K\bar{K}_1(1400)$ bound tate.  Besides the $P$-wave $\eta\eta^{'}$ decay model, the transitions from $\eta_1(1855)$ to final states
$K\bar{K}^{*}$,$\bar{K}^*K^*$, $f({1285})\eta$, $K\pi\bar{K}^{*}$, $K\bar{K}\rho$, and $K\bar{K}\omega$ are allowed.  Moreover, the radiative decay
width of $K\bar{K}_1(1400)$ molecular state are also evaluated.

This work is organized as follows. The theoretical
formalism is explained in Sec. II. The predicted partial
decay widths are presented in Sec. III, followed by a short summary in the last section.

\section{THEORETICAL FORMALISM}
\subsection{The decay $\eta_1(1855)\to\eta\eta^{'}$, $f_1(1285)\eta$, $\bar{K}^*K^*$,and $K\bar{K}^{*}$}
\begin{figure}[h!]
\begin{center}
\includegraphics[scale=0.96]{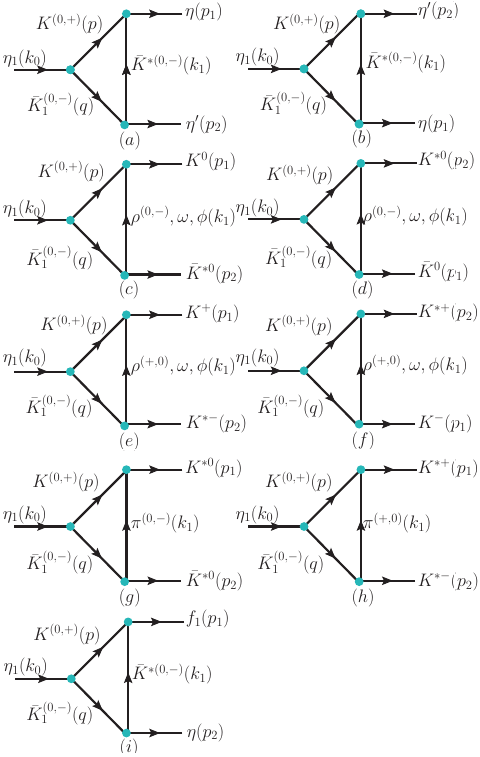}
\caption{Feynman diagrams for $\eta_1(1855)\to\eta\eta^{'}$, $f_1(1285)\eta$, and $K\bar{K}^{*}$ decay processes.
The contributions from $t$-channel $K^{*}$, $\rho$, $\pi$, $\phi$ and $\omega$ mesons  are considered.  We also show the definition
of the kinematical ($k_0$, $p$, $q$, $p_1$, $p_2$) that we use in the present calculation.}
\label{cc1}
\end{center}
\end{figure}

We first compute the two-body strong decay widths.  The relevant Feynman diagrams for the process $\eta_1(1855)\to\eta\eta^{'}$,
$f_1(1285)\eta$, $\bar{K}^*K^*$,and $K\bar{K}^{*}$ are shown in Fig.~\ref{cc1}.  The dominant mechanism of the vector meson exchanges
($\rho$, $\omega$, $\phi$, $K^*$) are considered.
To evaluate the diagrams shown in Fig.~\ref{cc1}, the following effective Lagrangians are needed~\cite{Wang:2019niy,Divotgey:2013jba,Dong:2020rgs}
\begin{align}
\mathcal{L}_{\eta_1 K_1 K } &=g_{\eta_1 K_1 K} \sum_{i=\bar{K}_1^0 K^0,\bar{K}_1^- K^+} {\cal{C}}_i \eta_1^{\mu}(x) \int d^4y \Phi (y^2)\nonumber\\
                            &\times{}K(x+ \omega_{K_{1\mu}} y )\bar{K}_{1 \mu }^{\dagger } (x - \omega_{K} y ) + h.c \label{eq1}\\
{\cal{L}}_{VPP}&=i \sqrt{2} G_V  \langle  [\partial_{\mu} P,P] V^{\mu} \rangle,\\
{\cal{L}}_{VVP}&= \frac{G'}{\sqrt{2}} \epsilon^{\mu\nu\alpha\beta} \langle \partial_{\mu} V_{\nu} \partial_{\alpha} V_{\beta} P \rangle \label{eq2} ,
\end{align}
where $G_V \approx 3.0 $ was estimated from the decay width of $\rho \to \pi \pi $. The coupling constant $G'=\frac{3g'^2}{4 \pi^2 f_{\pi} }$
is determined by the hidden gauge symmetry with  $g'=-\frac{{\cal{G}}_V m_{\rho} }{ \sqrt{2} f_{\pi}^2 }$, and ${\cal{G}}_V=55$ MeV, $f_{\pi}=93 $ MeV.
Since the $\eta_1(1855)$ carry quantum numbers $J^{PC}=1^{-+}$, the flavor function for a definite charge parity $C=+1$ can be determined from
Ref.~\cite{Liu:2013rxa}
\begin{align}
	|\bar{K}_1 K,I=0\rangle=\sqrt{\frac{1}{2}}(\bar{K}_1^0 K^0 + \bar{K}_1^- K^+),
\end{align}
with the following isospin assignments for $\bar{K}_1 $ and $ K $,
\begin{align}
	\left(
	\begin{array}{c}
		\bar{K}_1^0 \\ \bar{K}_1^-
	\end{array}
	\right)
	\sim
	\left(
	\begin{array}{c}
		|\frac{1}{2}, \frac{1}{2} \rangle \\	-|\frac{1}{2}, -\frac{1}{2} \rangle
	\end{array}
	\right)	,
	\left(
	\begin{array}{c}
		K^0 \\ K^+
	\end{array}
	\right)
	\sim
	\left(
	\begin{array}{c}
		|\frac{1}{2}, -\frac{1}{2} \rangle \\	|\frac{1}{2}, \frac{1}{2} \rangle
	\end{array}
	\right).	
\end{align}
That means the  ${\cal{C}}_i= 1/\sqrt{2}$, which is the product of the isospin factor and charge parity factor.p

$V_{\mu}$ and $P$ are the $SU(3)$ vector and pseudoscalar meson matrices, respectively, and $\langle\cdot\cdot\cdot\rangle$
denotes the trace in the flavor.  The meson matrices are~\cite{Wang:2019niy,Divotgey:2013jba,Dong:2020rgs}
\begin{align}
	P&=\left(
	\begin{array}{ccc}
		\frac{\eta_N + \pi^{0}}{\sqrt{2}} &   \pi^{+}   &  K^{+}   \\
		\pi^{-} & \frac{\eta_N - \pi^0}{\sqrt{2}} & K^{0} \\
		K^{-} & \bar{K}^{0} & \eta_S \\
	\end{array}
	\right) ,\\
	V&=\left(
	\begin{array}{ccc}
		\frac{\omega}{\sqrt{2}}+\frac{\rho^{0}}{\sqrt{2}} &   \rho^{+}   &  K^{*+}   \\
		\rho^{-} & \frac{\omega}{\sqrt{2}}-\frac{\rho^{0}}{\sqrt{2}} & K^{*0} \\
		K^{*-} & \bar{K}^{*0} & \phi \\
	\end{array}
	\right),
\end{align}
where $\eta = \eta_N \cos\varphi_P + \eta_S \sin \varphi_P $ and $\eta'= \eta_N sin\varphi_P + \eta_S cos \varphi_P$,
$\varphi_P=-41.46^{\circ}$, which implies the mixing of strange and non-strange isoscalar sector.

$g_{\eta_1 K_1 K}$ is the coupling constant and can be determined by the compositeness condition~\cite{Weinberg:1962hj,Salam:1962ap}.
It tells us that the $K\bar{K}^{*}$ molecular state must meet a relation, in which the renormalization constants of a bound state wave
function should be zero
\begin{align}
	Z_{\eta_1}=1-\frac{d\Sigma_{\eta_1}^T}{dk_0}|_{k_0=m_{\eta_1}}=0\label{ewqa}.
\end{align}
In the above equation, $\Sigma_{\eta_1}^T$ is the transverse part of the mass operator and
relates to its mass operator via the relation
\begin{align}
	\Sigma^{\mu\nu}_{\eta_1}(k_0)=(g_{\mu\nu}-\frac{k_0^{\mu}k_0^{\nu}}{k_0^2})\Sigma^{T}_{\eta_1}+\cdots.
\end{align}
\begin{figure}[h!]
	\begin{center}
		\includegraphics[scale=0.8]{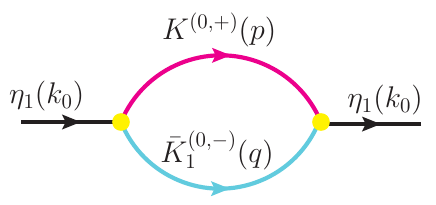}
		\caption{the self-energy of $\eta_1(1855)$ }
		\label{cc3}
	\end{center}
\end{figure}

\begin{figure}[h!]
\centering
\includegraphics[bb=00 100 700 400,clip,scale=0.55]{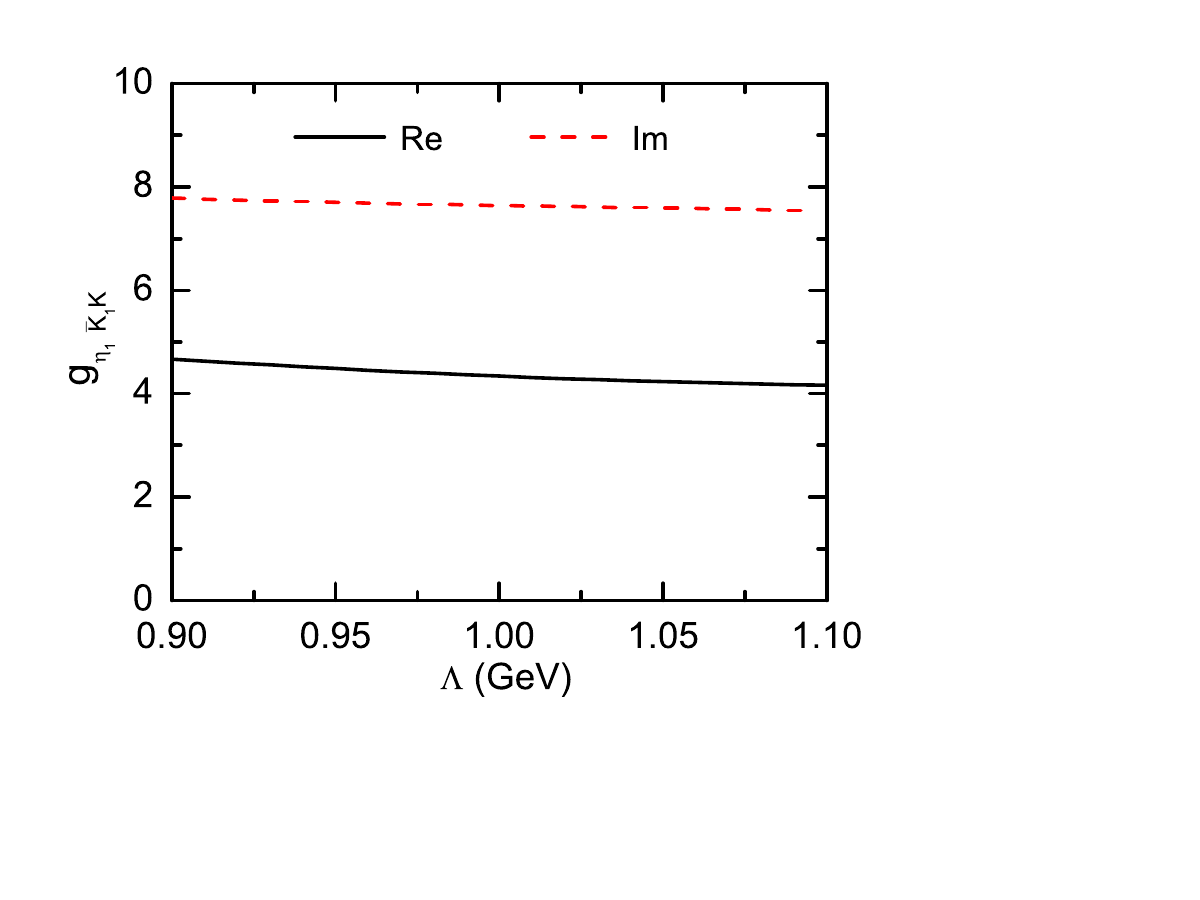}
\caption{(color online) The coupling constant of $g_{\eta_1 K_1 K}$ as a function of the parameter $\Lambda$.}
\label{fig.coupling}
\end{figure}

Consider the lowest order self-energy diagram that is shown in Fig.~\ref{cc3}, the mass operator $\Sigma^{\mu\nu}_{\eta_1}(k_0)$
can be determined by Eq.\ref{eq1}
\begin{align}
	\Sigma_{\eta_1}^{\mu\nu}&= i^2 g_{\eta_1 \bar{K}_1 K}^2 \sum_{i} C_i^2 \int \frac{d^4q}{(2\pi)^4}  \Phi^2[(p\omega_{\bar{K}_1}-q\omega_{K})^2]\nonumber\\
	&\times \frac{1}{p^2-m_{K}^2} \frac{-g^{\mu\nu}+ q^{\mu}q^{\nu}/m_{K_1}^2 }{q^2-m_{K_1}^2+im_{K_1}\Gamma_{K_1}}.
\end{align}
Where $\omega_{i}=m_i/(m_i+m_j)$ with $m_{i}$ is the masses of $K_1$ or $K$ meson.  $\Gamma_{K_1}=174$ MeV is the width of constituent meson $K_1$.
Obviously, the correlation function $\Phi (y^2)$ is introduced to stop the Feynmann diagrams ultraviolet infinite.  It always makes the amplitude
for the Feynman diagram shown in Fig.~\ref{cc3} to decrease fast to zero when $q$ varies from 0 to $+\infty$.  Here,
we would like to apply a widely used form, which is
\begin{align}
\Phi (p^2) \doteq \exp(-p_E^2/ \Lambda ^2),
\end{align}
where $p_E$ being the Euclidean Jacobi momentum.  And $\Lambda$ being the size parameter, which can only be determined from experimental data.
Fortunately, some stringent constraints for the $\Lambda$ value have been made by comparing with the experimental data and is determined to be
$\Lambda\simeq{}1.0$ GeV~\cite{Xiao:2019mvs,Huang:2020taj,Huang:2018wgr,Yang:2021pio,Dong:2009uf}.  Here the $\Lambda$ dependence of the coupling
constant $g_{\eta_1K\bar{K}_1}$ is confirmed and the results are plot in Fig.~\ref{fig.coupling}.

The effective Lagrangian describing vertices related to $K_1$ have been constructed in Refs~\cite{Dong:2020rgs,Dong:2022cuw,Divotgey:2013jba}.
And we give their explicit form as follow
\begin{align}
	\mathcal{L}_{K_{1A} K^* \eta }&=i \frac{3\sqrt{3}}{8} a \eta \bar{K}_{1A}^{0 \mu} K^{*0 \mu }\\
	 \mathcal{L}_{K_{1B} K^* \eta }&=- \frac{\sqrt{3}}{8} b \eta \bar{K}_{1B}^{0 \mu} K^{*0 \mu }\\
	\mathcal{L}_{K_{1A} K^* \eta' }&=i \frac{1}{\sqrt{3}} a \eta' \bar{K}_{1A}^{0 \mu} K^{*0 \mu }\\
	 \mathcal{L}_{K_{1B} K^* \eta' }&=- \frac{1}{3\sqrt{3}} b \eta' \bar{K}_{1B}^{0 \mu} K^{*0 \mu }\\
	 \mathcal{L}_{K_{1A} \rho K^0 }&=i \frac{1}{\sqrt{2}} a K^0 \bar{K}_{1A}^{0 \mu} \rho^{0 \mu }\\
	 \mathcal{L}_{K_{1B} \rho K^0 }&=- \frac{1}{\sqrt{2}} b \bar{K}^0 \bar{K}_{1B}^{0 \mu} \rho^{0 \mu }\\
     \mathcal{L}_{K_{1A} \omega K^0 }&=- i \frac{1}{\sqrt{2}} a K^0 \bar{K}_{1A}^{0 \mu} \omega^{ \mu }\\
	 \mathcal{L}_{K_{1B} \omega K^0 }&=\frac{1}{\sqrt{2}} b K^0 \bar{K}_{1B}^{0 \mu} \omega^{ \mu }\\
     \mathcal{L}_{K_{1A} K^{*0} \pi^0 }&=i \frac{1}{\sqrt{2}} a \pi^0 \bar{K}_{1A}^{0 \mu} K^{*0 \mu }\\
     \mathcal{L}_{K_{1B}  K^{*0} \pi^0 }&= -\frac{1}{\sqrt{2}} b \pi^0 \bar{K}_{1B}^{0 \mu} K^{*0 \mu }\\
     \mathcal{L}_{K_{1A} \phi K^0 }&=-i  a \phi^{\mu} K^0 \bar{K}_{1A}^{0 \mu} \\
      \mathcal{L}_{K_{1B} \phi K^0 }&= b \phi^{\mu} K^0 \bar{K}_{1B}^{0 \mu} \\
     \mathcal{L}_{K_{1A} K^{*0} \rho } &=\frac{1}{\sqrt{2}}a' \epsilon^{\alpha\beta\gamma\delta} \partial_{\alpha} \bar{K}_{1A \beta }^{0} \rho_{\gamma}^{0} K_{\delta}^{*0} \\
	\mathcal{L}_{K_{1B} K^{*0} \rho } &=i \frac{1}{\sqrt{2}}b' \epsilon^{\alpha\beta\gamma\delta} \partial_{\alpha} \bar{K}_{1B \beta }^{0} \rho_{\gamma}^{0} K_{\delta}^{*0} \\	
   	\mathcal{L}_{K_{1A} K^{*0} \omega } &=-\frac{1}{\sqrt{2}}a' \epsilon^{\alpha\beta\gamma\delta} \partial_{\alpha} \bar{K}_{1A \beta }^{0} \omega_{\gamma} K_{\delta}^{*0} \\
   \mathcal{L}_{K_{1B} K^{*0} \omega } &=- i\frac{1}{\sqrt{2}} b' \epsilon^{\alpha\beta\gamma\delta} \partial_{\alpha} \bar{K}_{1B \beta }^{0} \omega_{\gamma} K_{\delta}^{*0}	\\
   \mathcal{L}_{K_{1A} K^{*0} \phi } &=-a' \epsilon^{\alpha\beta\gamma\delta} \partial_{\alpha} \bar{K}_{1A \beta }^{0} \phi_{\gamma} K_{\delta}^{*0} \\
    \mathcal{L}_{K_{1B} K^{*0} \phi } &=- i b' \epsilon^{\alpha\beta\gamma\delta} \partial_{\alpha} \bar{K}_{1B \beta }^{0} \phi_{\gamma} K_{\delta}^{*0}
\end{align}
where $a=5.43$ and $b=-7.0$ are determined by the decay properties of $f_1(1420)$ and $b_1(1235) $ ~\cite{Dong:2022cuw,Divotgey:2013jba}.
$a'=1.15$ and $b'=-0.73$ are estimated by a quark model approach in Ref~\cite{Dong:2020rgs}.  The axial vector $K_{1A}$ and pseudovector $K_{1B}$ are two
parts of the physical state $K_1(1400)$.  And the mixing relation is parameterized as~\cite{Divotgey:2013jba}
\begin{align}
	| K_1(1400) \rangle = -i\sin\phi~ K_{1A} +  \cos\phi ~ K_{1B}
\end{align}
with the mixing angle $\phi=(56.4 \pm 4.3)^{\circ}$.

We also need the effective lagrangian for vertice $K^{*}K f_1$, which can be obtained from Ref.~\cite{Xie:2019iwz,Roca:2005nm}
\begin{align}
\mathcal{L}_{f_1K^*K}=-i{\cal{H}}_1g_{f_1}  \bar{K} K^*_{\mu} f_1^{\mu},
\end{align}
where $g_{f_1}$=7.555 GeV is obtained in the chiral unitary approach \cite{Roca:2005nm}.  ${\cal{H}}_1=-0.5$ and $0.5$ are for vertices $K\bar{K}^{*}f_1$ and $\bar{K}K^{*}f_1$, respectively.

Thus, we can obtain the following amplitudes for the decay
\begin{align}
\mathcal{M}_{a}&= i^5\frac{3}{4} \sqrt{\frac{3}{2}} \sum_{j}{\cal{C}}_j G_V g_{\eta_1 K_1 K } \int \frac{d^4 k_1 }{(2 \pi)^4 }(p_1^{\sigma} + p^{\sigma})\nonumber\\
           &\times\Phi[(p\omega_{K_{1}} -q \omega_{K})^2 ]\frac{- g^{\sigma \rho } + k_1^{\sigma} k_1^{\rho} / m_{K^{*}}^2 }{k_1^2 - m_{K^{*}}^2}\nonumber\\
           &\times{\cal{Y}}_{\mu\rho}\varepsilon^{\mu}(k_0) \frac{1}{p^2 - m_{K}^2  }, \\
\mathcal{M}_{b}&=-i^5\sqrt{\frac{2}{3}}\sum_{j}{\cal{C}}_j G_V g_{\eta_1 K_1 K } \int \frac{d^4 k_1 }{(2 \pi)^4 }(p_2^{\sigma} + p^{\sigma}  )\nonumber\\
          &\times\Phi[(p \omega_{K_{1}} -q \omega_{K})^2 ]\frac{- g^{\sigma \rho } + k_1^{\sigma} k_1^{\rho} / m_{K^{*}}^2 }{k_1^2 - m_{K^{*}}^2 } \nonumber\\
          &\times\frac{9}{8}{\cal{Y}}_{\mu\rho}\varepsilon^{\mu}(k_0) \frac{1}{p^2 - m_{K}^2  },\\
\mathcal{M}_{c}&=- i^5\sum_{j}{\cal{F}}_V{\cal{C}}_jG_V g_{\eta_1 K_1 K } \int \frac{d^4 k_1 }{(2 \pi)^4 }(p_1^{\sigma} + p^{\sigma}) \nonumber\\
    &\times\Phi [(p \omega_{K_{1}} -q \omega_{K})^2 ]\epsilon^{\alpha\nu\rho\lambda}\frac{- g^{\sigma \rho } + k_1^{\sigma} k_1^{\rho} / m_{V}^2 }{k_1^2 - m_{V}^2 }\nonumber\\
    &\times{\cal{W}}^{'}_{\mu\nu}q^{\alpha}\varepsilon^{\lambda \dagger }(p_2 ) \varepsilon^{\mu}(k_0)\frac{1}{p^2 - m_{K}^2  },\\
\mathcal{M}_{d}&=-i^5\sum_{j}\frac{{\cal{F}}_V}{2}{\cal{C}}_j G' g_{\eta_1 K_1 K } \int \frac{d^4 k_1 }{(2 \pi)^4 }\Phi [(p \omega_{K_{1}} -q \omega_{K})^2 ]\nonumber\\
      &\times\epsilon^{\alpha\sigma\beta\lambda } p_2^{\beta} k_1^{\alpha}\varepsilon^{\lambda \dagger }(p_2)\frac{-g^{\sigma \rho }+k_1^{\sigma}k_1^{\rho}/m_{V}^2 }{k_1^2 - m_{V}^2 }\nonumber\\
      &\times{\cal{W}}_{\mu\rho}\varepsilon^{\mu}(k_0) \frac{1}{p^2 - m_{K}^2},\\
\mathcal{M}_{e}&=-i^5\sum_{j}{\cal{P}}_V{\cal{C}}_j G_V g_{\eta_1 K_1 K } \int \frac{d^4 k_1 }{(2 \pi)^4 }\Phi [(p \omega_{K_{1}} -q \omega_{K})^2 ]\nonumber\\
    &\times\epsilon^{\alpha\nu\rho\lambda }(p_1^{\sigma} + p^{\sigma})\frac{- g^{\sigma \rho } + k_1^{\sigma} k_1^{\rho} / m_{V}^2 }{k_1^2 - m_{V}^2 }\nonumber\\
    &\times{\cal{W}}^{'}_{\mu\nu}q^{\alpha}\varepsilon^{\lambda \dagger }(p_2 ) \varepsilon^{\mu}(k_0)\frac{1}{p^2 - m_{K}^2},\\
\mathcal{M}_{f}&=-i^5\sum_j\frac{{\cal{P}}_V}{2}{\cal{C}}_j G'g_{\eta_1 K_1 K } \int \frac{d^4 k_1 }{(2 \pi)^4 }\Phi [(p \omega_{K_{1}} -q \omega_{K})^2 ]\nonumber\\
   &\times\epsilon^{\alpha\sigma\beta\lambda}p_2^{\beta}k_1^{\alpha}\varepsilon^{\lambda \dagger}(p_2)\frac{-g^{\sigma \rho }+k_1^{\sigma}k_1^{\rho}/m_{V}^2}{k_1^2-m_{V}^2 }\nonumber\\
   &\times{\cal{W}}_{\mu\rho}\varepsilon^{\mu}(k_0) \frac{1}{p^2 - m_{K}^2},\\
 \mathcal{M}_{g}&=
            -i^5\sum_j {\cal{U}}_P {\cal{C}}_j G_V g_{\eta_1 K_1 K } \int \frac{d^4 k_1 }{(2 \pi)^4 }(k_1^{\rho}-p^{\rho})\nonumber\\
            &\times  \Phi [(p \omega_{K_{1}} -q \omega_{K})^2 ] \varepsilon^{\rho \dagger}(p_1)   \frac{1}{k_1^2-m_{\pi}^2 }\varepsilon^{\lambda \dagger}(p_2)\nonumber\\
            &\times {\cal{W}}_{\mu\lambda} \varepsilon^{\mu}(k_0) \frac{1}{p^2 - m_{K}^2}\label{eq39} ,
 \end{align}
 \begin{align}
\mathcal{M}_{h}&=
            -i^5\sum_j {\cal{R}}_P {\cal{C}}_j G_V g_{\eta_1 K_1 K } \int \frac{d^4 k_1 }{(2 \pi)^4 }(k_1^{\rho}-p^{\rho})\nonumber\\
            &\times  \Phi [(p \omega_{K_{1}} -q \omega_{K})^2 ] \varepsilon^{\rho \dagger}(p_1)   \frac{1}{k_1^2-m_{\pi}^2 }\varepsilon^{\lambda \dagger}(p_2)\nonumber\\
            &\times {\cal{W}}_{\mu\lambda} \varepsilon^{\mu}(k_0) \frac{1}{p^2 - m_{K}^2}\label{eq40} ,\\
\mathcal{M}_{i}&=-i^4 \sum_j{\cal{C}}_jg_{f_1K^*K} g_{\eta_1 K_1 K } \int \frac{d^4 k_1 }{(2 \pi)^4 }\Phi [(p \omega_{K_{1}} -q \omega_{K})^2 ]\nonumber\\
               &\times\varepsilon^{\lambda \dagger }(p_1) \frac{- g^{\sigma \rho } + k_1^{\sigma} k_1^{\rho} / m_{K^{*}}^2 }{k_1^2 - m_{K^{*}}^2 }\frac{9}{8}{\cal{Y}}_{\mu\rho}\varepsilon^{\mu}(k_0) \frac{1}{p^2 - m_{K}^2},
\end{align}
with
\begin{align}
{\cal{Y}}_{\mu\rho}&=\bigg(\frac{1}{\sqrt{3} } a \sin \phi\frac{- g^{\mu \rho } + q^{\mu} q^{\rho} / m_{K_{1A}^{0}}^2 }{q^2 - m_{K_{1A}^{0}}^2 }\nonumber\\
           &-\frac{1}{3\sqrt{3} }b\cos\phi\frac{- g^{\mu \rho } + q^{\mu} q^{\rho} / m_{K_{1B}^{0}}^2 }{q^2 - m_{K_{1B}^{0}}^2 }\bigg),\\
{\cal{W}}^{(')}_{\mu\nu}&=\bigg(\frac{ 1 }{ \sqrt{2} } a^{(')} \sin \phi \frac{- g^{\mu \nu } + q^{\mu} q^{\nu } / m_{K_{1A}^{-}}^2 }{q^2 - m_{K_{1A}^{-}}^2 }\nonumber\\
	       &-\frac{1}{ \sqrt{2} } b^{(')} \cos \phi \frac{- g^{\mu \nu } + q^{\mu} q^{\nu} / m_{K_{1B}^{-}}^2 }{q^2 - m_{K_{1B}^{-}}^2 }\bigg).
\end{align}
In the above, ${\cal{F}}_{V}={\cal{P}}_{V}=2$ for $V=\rho^{\pm}$,$\phi$ while ${\cal{F}}_{V}={\cal{P}}_{V}=1$ for $V=\rho^{0}$,$\omega$.  Moreover,
${\cal{U}}_{P}={\cal{R}}_{P}=2$ and ${\cal{U}}_{P}={\cal{R}}_{P}=1$ for $P=\pi^{\pm}$ and $P=\pi^{0}$, respectively.

\subsection{Three-body decay}
\begin{figure}[h!]
	\begin{center}
		\includegraphics[scale=0.40]{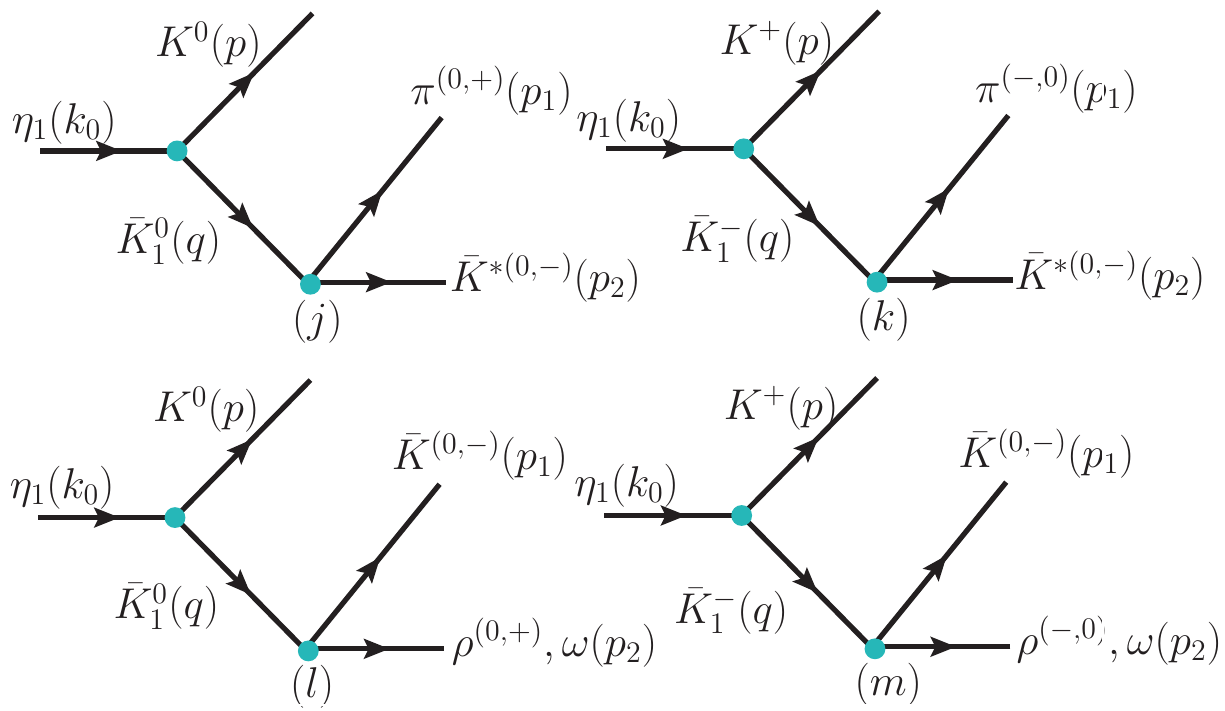}
		\caption{Three-body decay feynman diagrams of $\eta_1(1855)$. }
		\label{cc2}
	\end{center}
\end{figure}

In this section, we study the three-body decays of $\eta_1(1855)$ by assuming it as $K\bar{K}_1$ molecular.  Such assignment for the $K\bar{K}_1$ can make
$\eta_1(1855)$ decay to the final states directly through the simple tree diagram, rather than the loop diagram.  Because we think the two-body or three-body
decay modes of the multi-quark states through the tree diagram are usually the dominant ones.  Thus, we compute the decay of $K\bar{K}_1$ molecular into
$\pi{}K\bar{K}^{*}$, $K\bar{K}\rho$, and $K\bar{K}\omega$ three-body final states, and the relevant Feynman diagrams are shown in Fig.~\ref{cc2}.

With the lagrangians shown in last section, the amplitudes relates to the Fig.~\ref{cc2} can be simply got
\begin{align}
\mathcal{M}(\rightarrow K\pi\bar{K}^{*})&=i{\cal{I}}{\cal{C}}_jg_{\eta_1K_1K}\Phi[(p\omega_{K_{1}}-q\omega_{K^{0}})^2]{\cal{W}}_{\mu\nu}\nonumber\\
                       &\times\varepsilon^{\nu \dagger}(p_2 )\varepsilon^{\mu}(k_0),\\
\mathcal{M}(\rightarrow K\bar{K}\rho )&=i{\cal{I}}{\cal{C}}_j g_{\eta_1 K_1 K }  \Phi [(p \omega_{K_{1}} -q \omega_{K^{0}})^2 ]{\cal{W}}_{\mu\nu}\nonumber\\
                                      &\times\varepsilon^{\nu \dagger } (p_2 ) \varepsilon^{\mu} (k_0),\\
\mathcal{M}(\rightarrow K\bar{K}\omega )&=-i{\cal{C}}_j g_{\eta_1 K_1 K }  \Phi [(p \omega_{K_{1}} -q \omega_{K^{0}})^2 ]{\cal{W}}_{\mu\nu}\nonumber\\
                                      &\times\varepsilon^{\nu \dagger } (p_2 ) \varepsilon^{\mu} (k_0)
\end{align}
where the ${\cal{I}}$ is the isospin factor that relate to $\pi$ or $\rho$.

\subsection{Radiative decay}
\begin{figure}[h!]
\begin{center}
\includegraphics[scale=0.40]{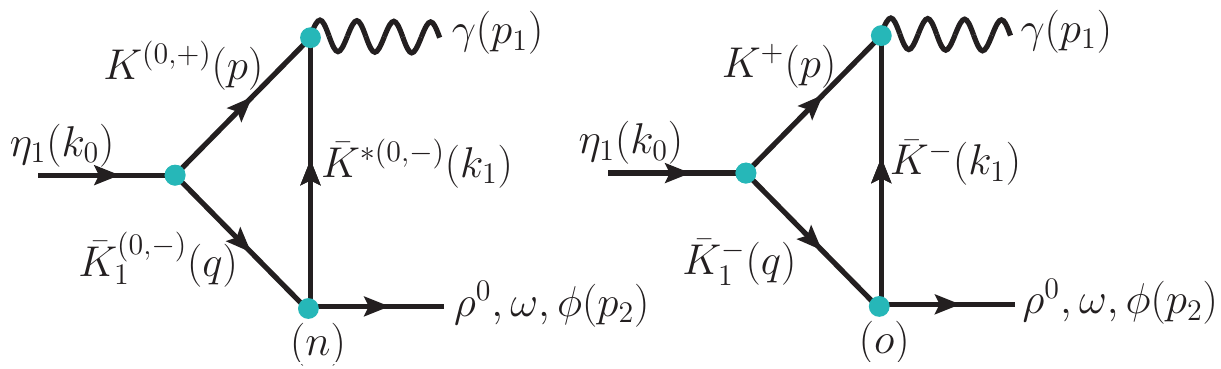}
\caption{The Feynman diagrams for the transition from $\eta_1(1855)$ to $\gamma{}\rho$, $\gamma{}\omega$, and $\gamma{}\phi$ under the $K\bar{K}_1$ assignment.}
\label{cc7}
\end{center}
\end{figure}

Here, we begin to compute the radiative decay. The interaction mechanisms for the processes $\eta_1(1855)\to{}\gamma{}V$ $(V=\omega,\rho,\phi)$
can be divided into two categories.  First includes mechanisms with the decay of $\eta_1(1855)$ to its molecular component $K\bar{K}_1$.  Then, the
decay $\eta_1(1855)\to{}\gamma{}V$ occur via the transitions $K\bar{K}_1\to \gamma{}V$ by considering $\bar{K}^{(*)}$ exchange.  The Feynman diagrams
are shown in Fig.~\ref{cc7}.  In this work, the transition from $\eta_1(1855)$ to $\gamma\bar{K}^0K$ through the tree diagram is ignored due to the
decay branching ratio of $K_{1}\to\gamma{}K^0$ is small and almost negligible.

To calculate the amplitudes of these diagrams, it is essential to know the Lagrangians for $\gamma{}KK^{(*)}$ vertexes.  Such Lagrangians have been
constructed in Refs.~\cite{Huang:2021ahp,Xie:2019iwz,Chen:2010re}, which are in the form of
\begin{align}
\mathcal{L}_{\gamma KK } &=i e A_{\mu} (K^- \partial^{\mu} K^+ - \partial^{\mu}  K^- K^+ ),\\
\mathcal{L}_{\gamma K K^* }&=g_{K^{*-} K^{-} \gamma } \epsilon^{\mu\nu\alpha\beta} \partial_{\mu} A_{\nu} \partial_{\alpha} K^{*-}_{\beta} K^-\nonumber\\
                           &+g_{K^{*0} K^{0} \gamma } \epsilon^{\mu\nu\alpha\beta} \partial_{\mu} A_{\nu} \partial_{\alpha} K^{*0}_{\beta} K^0 ,
\end{align}
where the coupling constants $g_{K^{*-}K^-\gamma }$=0.245$ GeV^{-1}$ and $g_{K^{*0}K^0\gamma }$=-0.388 $GeV^{-1} $ are determined
from the experimental widths $\Gamma(K^{*\pm} \to K^{\pm} \gamma )\approx 50$ keV and $\Gamma(K^{*0} \to K^{0} \gamma )\approx 116$ keV, respectively.
The signs of $g_{K^{*0}K^0\gamma }$ is fixed according to the quark model. $e =\sqrt{4\pi \alpha}$ with $\alpha$  being the fine-structure constant.

Then the radiative decay amplitudes corresponding to Fig.~\ref{cc7} can be obtained as follows
\begin{align}
\mathcal{M}_{n/o}=\mathcal{M}_{n/o}^{\eta}\varepsilon_{\eta}^{\dagger}(p_1)
\end{align}
with
\begin{align}
\mathcal{M}^{\eta}_n&=i^5{\cal{K}}_V{\cal{C}}_{j}g_{\eta_1 K_1 K } g_{\gamma K^{*0} K^0 }\int\frac{d^4 k_1 }{(2 \pi)^4 }\Phi [(p \omega_{K_{1}}-q\omega_{K})^2]\nonumber\\
             &\times\epsilon^{\alpha\eta\beta\sigma } \epsilon^{\omega\nu\lambda\rho}  p_1^{\alpha} k_1^{\beta} \frac{-g^{\sigma \rho }+k_1^{\sigma} k_1^{\rho}/m_{K^{*}}^2}
             {k_1^2 - m_{K^{*}}^2}\nonumber\\
             &\times{\cal{W}}^{'}_{\mu\nu}q^{\omega} \varepsilon^{\lambda \dagger }(p_2 ) \varepsilon^{\mu}(k_0)\frac{1}{p^2 - m_{K}^2},\\
\mathcal{M}^{\eta}_{o}&=i^5  {\cal{K}}_V {\cal{C}}_{j }eg_{\eta_1 K_1 K }\int \frac{d^4 k_1 }{(2 \pi)^4}\Phi [(p \omega_{K_{1}} -q \omega_{K})^2 ]\nonumber\\
               &\times(p^{\eta}-k_1^{\eta})  \frac{1 }{k_1^2 - m_{K}^2 }{\cal{W}}_{\mu\lambda}\varepsilon^{\lambda \dagger }(p_2) \varepsilon^{\mu}(k_0)\nonumber\\
               &\times\frac{1}{p^2 - m_{K}^2},                      	
\end{align}
where ${\cal{K}}_V=1$ for $V=\rho,\omega$ and ${\cal{K}}_V=\sqrt{2}$ for $V=\phi$.

We find that the relation $p_{1,\eta}(\mathcal{M}_l+\mathcal{M}_m)^{\eta}$ is not equal to zero.  It means the amplitudes that we obtained currently are not enough to satisfy the gauge invariance of the photon filed.  Therefore, the contact term ${\cal{M}}_{c}$ that has been used in Refs.~\cite{Huang:2021ahp,Zhu:2020lza} must be included.  The detailed calculations can be found in these two works and are not shown here.

Once the amplitudes are calculated, we can obtain the partial decay widths
\begin{align}
	d \Gamma(\eta_1 \to \eta\eta' , \bar{K} K^*,\gamma{}V )&=\frac{1}{2J+1}\frac{1}{32\pi^2}\frac{|\boldmath
		                         P_1|}{m_{\eta_1}^2}|\bar{{\cal{M}}}|^2 d\Omega  \\
	                     ~
	d\Gamma(\eta_1\to\bar{K}^{*}K\pi,\bar{K} K \rho,\bar{K} K  \omega  )&=
	                            \frac{1}{2J+1}\frac{1}{(2\pi)^5}\frac{1}{16M^2}\bar{|{\cal{M}}|^2}|\vec{p}^{*
		                        }_2|\nonumber\\
	                            &\times{}|\vec{p}|dm_{12}d\Omega^{*}_{p_2}d\Omega_{p},
\end{align}
where  $J$ is the total angular momentum of the $\eta_1$,  $|\vec{p}_1|$ is the three-momenta of the decay products in the center
of mass frame, the overline indicates the sum over the polarization vectors of the final hadrons.  The ($\vec{p}^{*}_2,\Omega^{*}_{p_2}$)
is the momentum and angle of the particle $\bar{K}^{*}$   or $\rho$,$\omega$ in the rest frame of $\bar{K}^*$ and $\pi$ or $\bar{K} $ and
$\rho$,$\omega$ ,  and $\Omega_{p}$ is the angle of the $K$ in the rest frame of the decaying particle.  The $m_{12}$ is the invariant mass
for $\pi$ and $K$ or $\bar{K} $ and $\rho$,$\omega$ ,with $m_{1}+m_{2}\leq{}m_{12}\leq{}M-m_{K}$.

\section{RESULTS And Discussions}
In this work, we compute the decay models of $\eta_1(1855)$.  It mainly decays to $\eta\eta^{'}$, $K\bar{K}^{*}$,$\bar{K}^*K^*$,
$f_1(1285)\eta$, $K\bar{K}^{*}\pi$, $K\bar{K}\rho$, and $K\bar{K}\omega$ by accepting that $\eta_1(1855)$ is a $K\bar{K}_1$
molecule.  In addition, its radiation decay widths are also studied and can be better used to test the molecular nature of the
$\eta_1(1855)$.

In Fig.~\ref{fig.coupling}, the coupling constant $g_{\eta_1 K_1 K}$ as a function of the $\Lambda$ are shown.  We can find that
the coupling constant $g_{\eta_1 K_1 K}$ decreases continuously but relatively slowly with the increase of $\Lambda$.
Varying the parameter $\Lambda$ from 0.9 to 1.1 GeV, the real and an imaginary component of the $g_{\eta_1 K_1 K}$
runs from 4.16 to 4.67 and 7.53 to 7.78, respectively,  which is not very sensitive to the model parameter $\Lambda$.
In particular, the Re$[g_{\eta_1 K_1 K}]=4.34$ and Im$[g_{\eta_1 K_1 K}]=7.64$ when we adopt the value $\Lambda=1.0$ GeV.

\begin{figure}[h!]
\begin{center}
\includegraphics[scale=0.35]{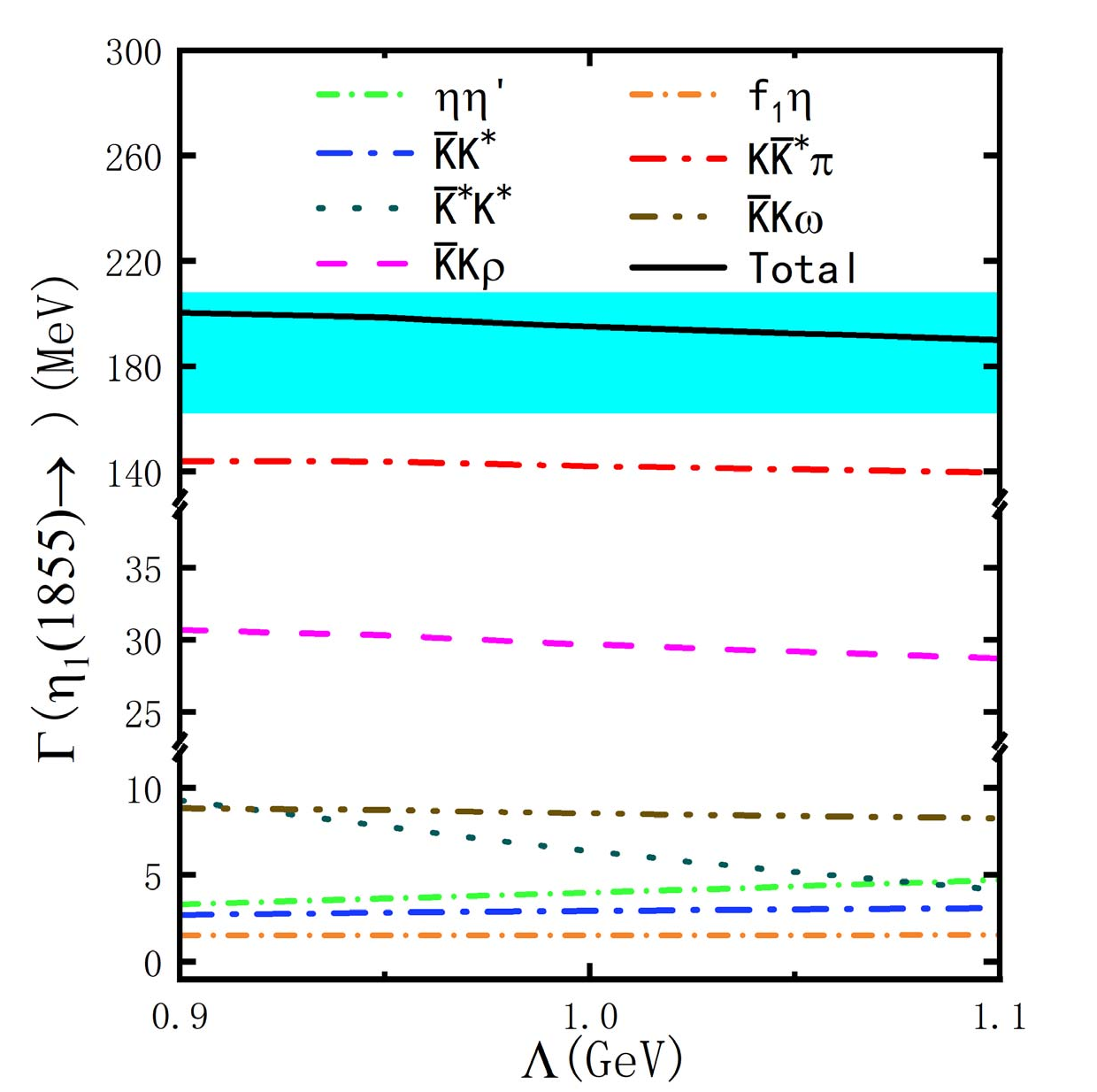}
\caption{Decay widths of $\eta_1(1855)$ with black dash line for total decay width, red dash dot line for $K\bar{K}^*\pi$ channel,
purple dash line for $\bar{K}K\rho$ channel, green dash dot line for $\eta \eta'$ channel, blue dash dot line for $\bar{K}K^*$ channel,
and brown dash dot dot line for $\bar{K}K\omega$ channel.  The cyan bands denote the experimental total width~\cite{Ablikim:2022zze,Ablikim:2022glj}.}\label{cc5}
\end{center}
\end{figure}

With the Lagrangians and the coupling constant $g_{\eta_1K_1K}$ obtained, the decay widths versus the
model parameter $\Lambda$ are calculated and presented in Fig.~\ref{cc5}.  Among the two-body decay models, we find that the
partial decay width of $K^{*}\bar{K}^{*}$ channel is the largest, $\eta\eta^{'}$ and $K\bar{K}^{*}$ channels are intermediate,
and $f_1(1285)\eta$ channel is the smallest.  Such small $f_1(1285)\eta$ decay width can be easily understood due to the smallest
phase space compared with the other two channels.  The decay mode $\eta\eta^{'}$ is suppressed since it is a $P$-wave decay,
rather than $S$-wave because the lowest angular momentum gives the dominant contribution.

However, the partial decay width of the $\eta_1(1855)\to{}K^{*}\bar{K}^{*}$ is much smaller than that of the $K\bar{K}^{*}\pi$ three-body decay.
Detailed numerical results are listed in Tab.~\ref{tab1}.  We find that the $K\bar{K}^{*}\pi$ three-body decay width is estimated to be 142.07
MeV at $\Lambda=1.0$ GeV, which is approximately twenty-two times bigger than that of the $\eta_1(1855)\to{}K\bar{K}^{*}$ decay.  It is because
the $K\bar{K}_1$ assignment for $\eta_1(1855)$ can decay to the final state $K\bar{K}^{*}\pi$ by occurring at the tree level, rather than via the
triangle diagrams for the $K\bar{K}^{*}$ decay.  And the two-body or three-body decay modes of the multi-quark states through the tree
diagram are usually the dominant ones.  We also find that the partial decay widths of $K\bar{K}\rho$ and $K\bar{K}\omega$ channels are
smaller than that of $K\bar{K}^{*}\pi$ channel due to the relatively small space phase.  The more important reason for this is that $K_1(1400)$
has the largest $\pi{}K^{*}$ decay branching ratio, while the decay widths of $K_1(1400)\to{}K\rho$ and $K_1(1400)\to{}K\omega$ are quite small,
about $3\%$ and $1\%$ of $K_1(1400)$ experiment width, respectively~\cite{Zyla:2020zbs}.
\begin{table}[h!]
\begin{center}
\caption{Partial decay widths of $\eta_1(1855)\to\eta\eta^{'}$, $K\bar{K}^{*}$, $K^{*}\bar{K}^{*}$, $f_1(1285)\eta$, $K\bar{K}^{*}\pi$, $K\bar{K}\rho$,
$K\bar{K}\omega$, and the total decay width with $\Lambda=1.0$ GeV. Error reflects variation of the $\Lambda$ in from 0.9 to 1.1 GeV.}\label{tab1}
\begin{tabular}{cc|cc} 		 	
  \hline
  ~~\textbf{Strong} ~~~&~~~ \textbf{Width(MeV)}      ~~~&~~~ \textbf{Radiative} ~~&~~ \textbf{Width(KeV)}      \\\hline		
  ~~$\eta \eta'$    ~~~&~~~ $3.98^{+0.73}_{-0.68}$   ~~~&~~~ $\gamma\rho$       ~~&~~ $12.63^{+0.14}_{-0.34}$  \\
  ~~$K\bar{K}^*$    ~~~&~~~ $2.92^{+0.16}_{-0.22}$  ~~~&~~~ $\gamma\omega$     ~~&~~ $12.50^{+0.14}_{-0.34}$  \\
  ~~$\bar{K}^*K^*$     ~~~&~~~ $6.36^{+2.93}_{-2.21}$   ~~~&~~~ $\gamma\phi$       ~~&~~ $17.67^{+0.38}_{-0.62}$ \\
  ~~$f_1 \eta $     ~~~&~~~ $1.51^{+0.03}_{-0.00}$      ~~~&~~~                    ~~&~~                         \\
  ~~$K\bar{K}\rho$  ~~~&~~~ $29.68^{+1.00}_{-0.98}$  ~~~&~~~                    ~~&~~                         \\
  ~~$K\bar{K}\omega$~~~&~~~ $8.53^{+0.30}_{-0.29}$   ~~~&~~~                    ~~&~~                         \\
  ~~$K\bar{K}^*\pi$ ~~~&~~~ $142.07^{+8.42}_{-1.65}$ ~~~&~~~                    ~~&~~                         \\
  \hline
  ~~Total          ~~~&~~~ $195.06^{+5.18}_{-5.13}$ ~~~&~~~  Total             ~~&~~ $42.80^{+0.65}_{-1.29}$ \\
  ~~Exp.~\cite{Ablikim:2022zze,Ablikim:2022glj}~~~&~~~ $188\pm18^{+3}_{-8}$ ~~~&~~~              ~~&~~  \\
  \hline
  \end{tabular}
 \end{center}
\end{table}

We also find that the total decay width is predicted to be about $195.06^{+5.18}_{-5.13}$ MeV, which can be confronted with the
experimental data.  It means that the total experimental decay width can be well reproduced, which provides direct evidence that
the observed $\eta_1(1855)$ is an $S$-wave $K\bar{K}_1$ molecular state.  And many works~\cite{Xiao:2019mvs,Huang:2020taj,Huang:2018wgr,Yang:2021pio,Dong:2009uf}
alway tell us that for an $S$-wave molecule the coupling strength of a bound state to its components is insensitive to the
$\Lambda$, which reflects the inner structure of the molecule.

As the same with the weak dependence of the coupling constant on the parameter $\Lambda$, the strong decay models are not also very
sensitive to the model parameter $\Lambda$. The results in Fig.~\ref{cc5} and Tab.~\ref{tab1} display that the partial decay
widths of $\eta_1(1855)\to\eta\eta^{'}$, $K\bar{K}^{*}$, and $f_1(1285)\eta$ increase slowly from 3.30 to 4.71 MeV, 2.70 to 3.08
MeV, 1.51 to 1.54 MeV, respectively.  While the decay widths of the transition from  $\eta_1(1855)$ to $\bar{K}^*K^*$,$K\bar{K}^{*}\pi$,$K\bar{K}\rho$,
and $K\bar{K}\omega$ monotonously decreases with increasing $\Lambda$, variation from 9.29 to 4.15 MeV,150.49 to 140.42 MeV, $30.68$ to 28.70 MeV,
and 8.83 to 8.24 MeV, respectively.  Since the three-body transition is the main decay channel, the dependence of the total decay
width on $\Lambda$  is the same as that of the three-body decay width and can be found in Fig.~\ref{cc5}.

It is worth noting that the radiation decay width is not included in the current total decay width.  The main reason for this is that the
radiation decay width has the transition strength often in the keV regime and is significantly lower than their strong counterparts.  However,
radiative decay is a better way to reveal the inner structure of an exotic state.  This is because the quark can interact directly with photons,
which is different from the $\eta_1(1855)$ interacting with photons through its molecular component $K\bar{K_1}$.

\begin{figure}[h!]
\begin{center}
\includegraphics[bb=-100 20 850 550, clip,scale=0.28]{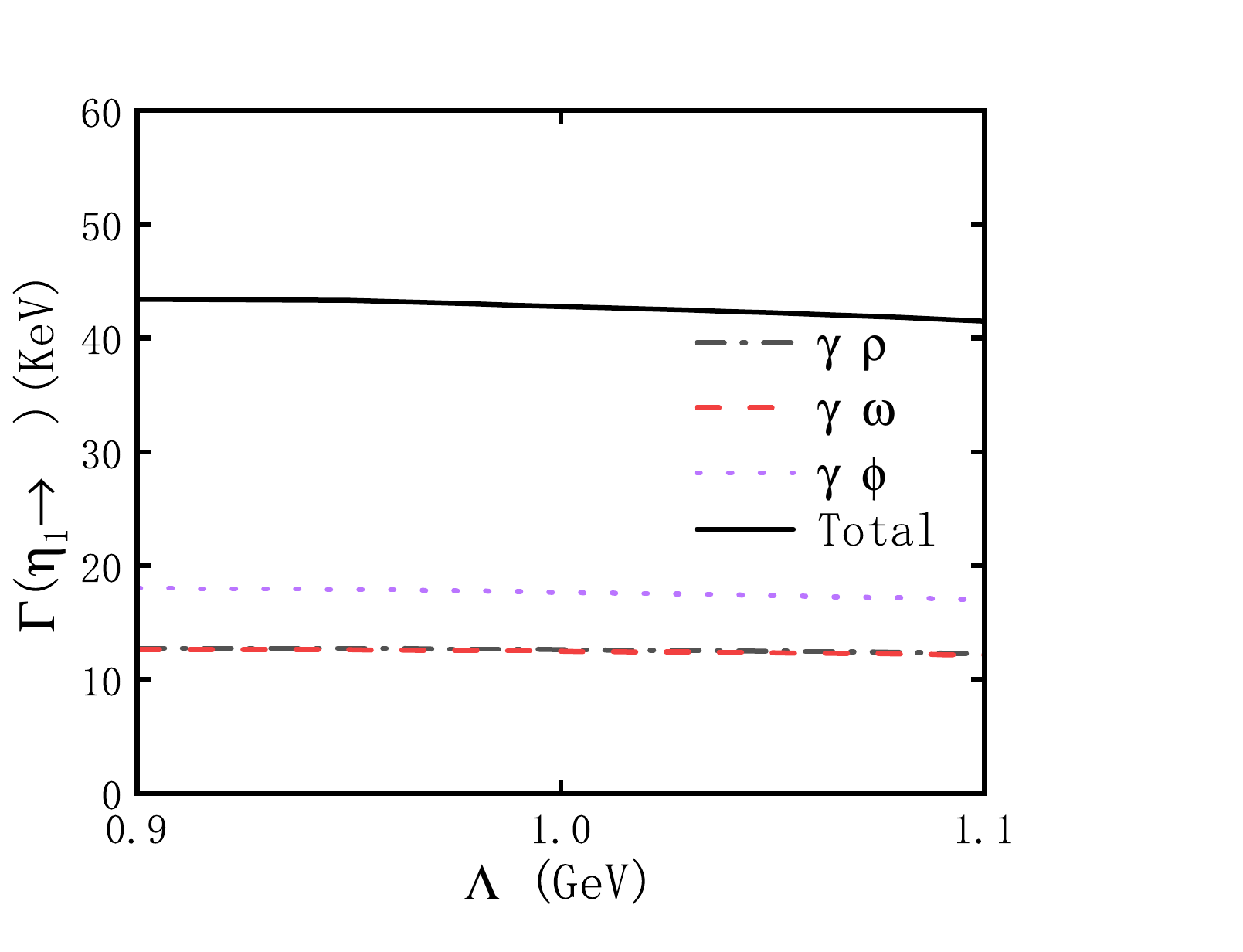}
\caption{(color online) Partial decay widths of the $\eta_1(1855)\to \gamma{}\rho$, $\eta_1(1855)\to \gamma{}\omega$, and $\eta_1(1855)\to
\gamma{}\phi$.}\label{cc15}
\end{center}
\end{figure}

The dependence of the corresponding radiative decay widths $\eta_1(1855)\to \gamma{}\rho$, $\eta_1(1855)\to \gamma{}\omega$, and
$\eta_1(1855)\to\gamma{}\phi$ on $\Lambda$ are depicted in Fig.~\ref{cc15}.  Similar to the three-body decay width, the radiative decay
widths gradually decrease with the increase of the $\Lambda$.  And the dependency on $\Lambda$ is also weak.  We also find that the
$\eta_1(1855)\to{}\gamma{}\phi$ is the main decay channel, while the $\eta_1(1855)\to \gamma{}\rho$ and $\eta_1(1855)\to \gamma{}\omega$
proved a small contribution.  A possible explain for this is that the $\eta_1(1855)\to \gamma{}\rho$ and $\eta_1(1855)\to \gamma{}\omega$
involve the creation or annihilation of two quark pairs, which are usually strongly suppressed.

Detailed numerical results for $\eta_1(1855)\to \gamma{}\rho$, $\eta_1(1855)\to \gamma{}\omega$, and $\eta_1(1855)\to\gamma{}\phi$ are
also presented in Table.~\ref{tab1}.  The center value corresponds to $\Lambda = 1.0$ GeV and error reflects variation of the $\Lambda$
in from 0.9 to 1.1 GeV.  Our calculation indicates the partial widths for $\eta_1(1855)\to \gamma{}\rho$, $\eta_1(1855)\to \gamma{}\omega$,
and $\eta_1(1855)\to\gamma{}\phi$ are very small and are evaluated to be $12.63^{+0.14}_{-0.34}$ KeV, $12.50^{+0.14}_{-0.34}$ KeV, and
$17.67^{+0.38}_{-0.62}$ KeV, respectively.

At present, our calculation supports the $\eta_1(1855)$ is an $S$-wave $K\bar{K}_1$ molecular state.  We find that the $K\bar{K}^{*}\pi$
three-body decay provides the dominant contribution, not the $\eta\eta^{'}$ channel found in the experiment.  The experimental
measurements for such a strong decay process could be crucial to test the nature of the $\eta_1(1855)$.  Because the $\eta\eta^{'}$
decay channel play the dominant role when the $\eta_1(1855)$ is explained as the $\bar{s}sg$ hybrid meson~\cite{Chen:2022qpd}.
The partial width for $\eta_1(1855)\to\gamma{}\phi$ can reach up to $17.67^{+0.38}_{-0.62}$ KeV, which can be detected in many experiment.
Such as the LHCb experiment.  It can also help us to distinguish whether the $\eta_1(1855)$ is a $K\bar{K}_1$ molecular or $\bar{s}sg$ hybrid
meson.

\section{summary}
In this work, the newly observed exotic state $\eta_1(1855)$ is investigated in $K\bar{K}_1(1400)$ molecular scenario.  The coupling between
the $\eta_1(1855)$ and its component $K\bar{K}_1$ is computed by the Weinberg compositeness condition.  With the help of an effective
Lagrangian approach,  the two-body and three-body strong decays of $\eta_1(1855)$ are evaluated through a triangle diagram and tree-level
diagram, respectively.   The experimental analysis can be reproduced by our theoretical calculations with that the numerical results are
shown in Fig.~\ref{cc5} and  Table.~\ref{tab1}.   The decay channel $\eta_1(1855)\to K\bar{K}^*\pi$ provides the dominant contribution,
instead of $\eta\eta'$ channel, which is observed in BESIII Collaboration.  Moreover, radiative decay of $\eta_1(1855)$ into $\gamma\phi$,
$\gamma\omega$, and $\gamma\rho$ are also studied in this work.   The dominant channel is $\gamma \phi$, and the partial width can reach
up to 17.67 KeV.   Further experiments in LHCb and BESII will support a crucial test for our investigation.

Although the studies of Ref.~\cite{Dong:2022cuw} and our work seem to indicate that the $\eta_1(1855)$ is a pure molecular state,  we cannot fully
exclude other possible explanations such as the $\bar{s}sg$ hybrid meson~\cite{Chen:2022qpd} or compact multiquark state (as long as quantum
numbers allow, it might well be the case).  Searching for the radiative and strong decay model of $\eta_1(1855)$ can help us to understand its
nature.  This is because the different partial decay widths rely on the structure assignments of $\eta_1(1855)$.

\section*{Acknowledgements}
Yin Huang acknowledges the YST Program of the APCTP.
This work was supported by the National Natural Science Foundation
of China under Grant No.12104076, the Science and Technology
Research Program of Chongqing Municipal Education Commission
(Grant No. KJQN201800510), and the Opened Fund
of the State Key Laboratory on Integrated Optoelectronics
(GrantNo. IOSKL2017KF19).  We also want to thanks
the support from the Development and Exchange Platform for
the Theoretic Physics of Southwest Jiaotong University under
Grants No.11947404 and No.12047576,  and the National Natural
Science Foundation of China under Grant No.12005177.


\end{document}